\documentstyle[preprint,prl,aps]{revtex}
\tighten
\begin{document}

\draft

\title{Calculation of Effective Coulomb Interaction for $Pr^{3+}$, $U^{4+}$,
and $UPt_3$}

\author{M.R. Norman}
\address{Materials Science Division, Argonne National Laboratory,
Argonne, IL 60439}

\date{\today}

\maketitle

\begin{abstract}

In this paper, the Slater integrals for a screened Coulomb interaction of the
the Yukawa form are calculated and by fitting the Thomas-Fermi
wavevector, good agreement is obtained with experiment for the multiplet
spectra of $Pr^{3+}$ and $U^{4+}$ ions.  Moreover, a
predicted multiplet spectrum for the heavy fermion superconductor $UPt_3$ is
shown with a calculated Coulomb U of 1.6 eV.  These effective
Coulomb interactions, which are quite simple to calculate, should be useful
inputs to further many-body calculations in correlated electron metals.

\end{abstract}


\narrowtext

Calculations of effective Coulomb interactions are important
not only in atomic spectroscopy but also in condensed matter physics where
the problem of correlated electron metals has received renewed attention
with the discovery of heavy fermion and high temperature superconductivity.
Even for an isolated atom, these calculations are difficult given the large
number of possible configurations which must be considered and in the case of
heavy ions, the added complication of taking into account relativistic effects.
Only recently have detailed configuration interaction (CI)
calculations been performed on $Pr^{3+}$,\cite{cai}
the prototype ion usually considered when trying to understand f electron
correlations.\cite{mr}  No similar calculations exist for $U^{4+}$, the
actinide analogue of $Pr^{3+}$.  The latter ion is of significance in
condensed matter physics since uranium heavy fermion metals are lattices of
such ions.\cite{gang}  In fact, recent studies of the $f^1$-$f^2$ Anderson
lattice model show that atomic correlations can have a profound impact on the
mean-field quasiparticle bands.\cite{trees}  Moreover, it has been proposed
that such correlations also appear in the residual quasiparticle-quasiparticle
interactions and are responsible for heavy fermion
superconductivity.\cite{vdm,norm}
If anything, a proper understanding of these atomic correlations will
be necessary to achieve a complete picture of heavy fermion physics and
probably copper oxides as well.  In this paper, a simple method for
calculating these correlation effects is presented.  The screened
interaction is taken to be of the Yukawa form and is evaluated using
relativistic wavefunctions from a Hartree-Fock program.\cite{mchf}  The
calculations are then fit with a single parameter (the Thomas-Fermi wavevector)
to the observed multiplet spectra of $Pr^{3+}$ and $U^{4+}$.  Moreover, results
of recent high energy neutron scattering data are used to make a
prediction of the multiplet spectrum for the heavy fermion superconductor
$UPt_3$.

The standard method for interpreting multiplet spectra in heavy ions is to
perform a least squares fit of the spectra using a Hamiltonian with adjustable
coefficients.\cite{gold1}  The main terms for $f^n$ spectra
are the effective Slater integrals $F^L$ (L=0,2,4,6) and the spin-orbit
splitting.  Smaller terms corresponding to odd L due to configuration
interaction plus other magnetic terms, which are necessary to get an exact fit
to the spectra, are ignored here.  These effective Slater integrals are
considerably smaller than their Hartree-Fock values.  This is as expected
since the effect of correlations is to screen these values (in CI language,
the $f$ electron is distributed in other configurations, thus reducing the
wavefunction overlap and decreasing the value of these integrals).  It is
difficult, though, to calculate these effective integrals from first
principles.
Morrison and Rajnak have shown that the corrections from many-body perturbation
theory have a slow convergence for $Pr^{3+}$.\cite{mr}  Recently, Cai et
al\cite{cai} have performed
detailed configuration interaction calculations for $Pr^{3+}$ including up
to 1708 configurations for each J state.  The agreement with experiment is
significantly improved over Hartree-Fock, but there are still some
discrepancies.

Here, an approach intermediate between the fitting procedure and
that of the ab initio calculations is taken.  The reason for the problems
mentioned above is that both ab initio methods are rather inefficient in
calculating these effects.  In solid state physics, one has to sum infinite
series to describe screening effects properly, so the
lack of convergence at third order in the above many-body perturbation
calculation is not too surprising for an atom with a large number of electrons
like $Pr$.  As for CI, wavefunction expansions of this sort have very slow
convergence except for small atoms when most of the correlation is coming from
near degeneracy
effects such as $Be$.  Again, these problems are more severe when large
numbers of electrons are involved.  The most efficient way of dealing with
this, then, is to assume a screened interaction from the outset.  In
particular, in this paper, a Yukawa form is used
\begin{equation}
V(\vec r_1, \vec r_2) = \frac{e^{-\lambda|\vec r_1 - \vec r_2|}}{|\vec r_1 -
\vec r_2|}
\end{equation}
The work of this paper is still semi-phenomenological in that $\lambda$,
the Thomas-Fermi wavevector, is treated as an adjustable constant.  Ab initio
calculations of it would suffer the same slow convergence problems as mentioned
above for the many-body perturbation calculation, but at least our ignorance
has been transfered to just one parameter as opposed to several of them as in
the least squares fitting to the experimental spectra discussed above.  In
solids, reducing the fitting parameter to one number is highly desirable since
the detailed information available in atomic spectroscopy is often lost due
to hybridization effects.

The Slater integrals of the Yukawa model are easily found.  Just as the bare
Coulomb interaction is expandable in powers of $r$, the screened
interaction is expandable in spherical functions\cite{math}
\begin{equation}
V(\vec r_1, \vec r_2) = -\lambda \sum_L (2L+1) j_L(i \lambda r_<)
h_L^{(1)}(i \lambda r_>) P_L(\cos\theta)
\end{equation}
where $j_L$ is a spherical Bessel function, $h_L^{(1)}$ is a spherical Hankel
function of the first kind, and $P_L$ is a Legendre polynomial, with $\theta$
the angle between $\vec r_1$ and $\vec r_2$ and $r_<$ ($r_>$) the lesser
(greater) of the two.  These spherical functions were determined using routines
in Numerical Recipes.\cite{nr}  The wavefunctions used in the Slater integrals
were determined from relativistic Hartree-Fock calculations.\cite{mchf}  For
convenience, a single integral is defined by taking a weighted contribution
of the two spin-orbit wavefunctions, that is $\phi_f^2$ in the
non-relativistic Slater integrals is replaced by
$\frac{3}{7}\phi_{5/2}^2 + \frac{4}{7}\phi_{7/2}^2$.

In Fig. 1, the evolution of these Slater integrals versus $\lambda$ for the
$U^{4+}$ case is shown ($\lambda$=0 is the Hartree-Fock result).
The L=0 integral is strongly screened whereas the higher L integrals are
less affected.  This is to be expected since the former is a charge fluctuation
integral whereas the latter are shape fluctuation integrals.  For all L,
the larger the L, the smaller the screening effect, as expected.

The matrix elements of the secular matrix for the $f^2$ ion case are listed
by Goldschmidt.\cite{gold2}  There are 13 eigenvalues (3 triplets, 4 singlets,
each of the triplets being split by spin-orbit).  The
spin-orbit splitting determined from the relativistic Hartree-Fock orbitals
matches the experimental estimate\cite{gold1} for $Pr^{3+}$ to within 0.1\%.
This is as expected since the spin-orbit interaction is a one-body force.
Implications from the fits that this interaction is screened was incorrect
because non-relativistic estimates had been used for the Hartree-Fock value.
A slight discrepancy was found in the $U^{4+}$ case in that the Hartree-Fock
value underestimated the experimental fit by about 2\%.  This is probably due
to the
strong intermediate coupling (LS versus jj) nature of the $U$ ion.  For the
purposes of this paper, the calculated Hartree-Fock values of $\xi$ are used
(where 7/2 $\xi$ is equal to the splitting of the 5/2 and 7/2 levels) as
opposed to the "experimental" ones.  These values are 0.0990 eV for $Pr^{3+}$
and 0.2387 eV for $U^{4+}$.  The secular matrix is then diagonalized using
Slater integrals for a particular value of $\lambda$.  The energy
of all levels will be refered to that of the ground state ($^3H_4$).

Results are shown for $Pr^{3+}$ in Fig. 2 with Slater integrals listed in
Table 1.  An RMS error of 82 meV is obtained for $\lambda$=2.0.  If the
highest level ($^1S_0$) is ignored, an RMS error of 40 meV is obtained for
$\lambda$=1.9.  This is to be compared to the RMS error of the CI
calculation\cite{cai} which is 163 meV.  An interesting point is that
the ordering of the levels agrees with experiment, therefore questions
raised\cite{cai} about the assignment of the $^1I_6$ level are probably not
valid.  Another point is that the values of the L=2 and L=4 Slater
integrals agree with the fitted (experimental) values.  On the other hand,
the L=6 one is substantially larger.  The quoted
experimental value may be somewhat misleading, as it is more screened
over Hartree-Fock than than the L=4 one, whereas one would conceptually expect
that higher L integrals are less screened than lower L ones (as is indeed
reproduced here).  This leads to the possibility that there is another least
squares fit to the data which more closely matches the current predictions.
Such a fit would presumably have different values of the odd L interactions
than listed in the published fits.  This issue will be discussed in more detail
in the next paragraph.  Finally, the calculated
L=0 Slater integral (the Coulomb repulsion, U) is close to the experimental
estimates of 5.3 eV\cite{brewer} and 5.5 eV\cite{lang}.  This is quite
interesting since the fit of $\lambda$ does not involve U.  This suggests that
the current method provides a trivial and independent
way of estimating the Coulomb U.  These estimates of U are important
inputs to many-body calculations in solids.

In Fig. 3 and Table 1, results are shown for $U^{4+}$.
For $\lambda$=1.6, an RMS error
of 84 meV is obtained.  If the highest level ($^1S_0$) is ignored, an RMS
error of 40 meV is obtained for $\lambda$=1.5.  These deviations are
identical to that found for $Pr^{3+}$, with $\lambda$ smaller in this case
since the 5f orbitals are more delocalized than the 4f ones.  The level
ordering agrees with experiment except for an interchange of $^1D_2$
and $^1G_4$.  The level ordering in this energy range (including $^3P_0$)
is rather sensitive to $\lambda$, and this minor problem is
almost certainly due to the intermediate coupling nature of this ion.
This has been checked by directly replacing the Slater integral routines in the
relativistic Hartree-Fock code by the Yukawa ones (there is a large number
of Slater integrals due to the various combinations of the two spin-orbit
radial functions).  In this case, the ordering of the two levels comes out
correct, but at the expense of having the $^3P_0$ lower than these two, as
opposed to higher as in experiment.  In fact, the RMS errors
are larger with this approach; in particular, the splittings within an LS
term are worse than with the other approach (these splittings have a weak
dependence on $\lambda$).  This again points to the intermediate coupling
nature of the ion as the problem.  Finally, as in $Pr^{3+}$, the
predicted L=6 Slater integral is much less strongly screened than implied
by the experimental least squares fits.\cite{vd}  On the other hand, the
value reported by Goldschmidt\cite{gold2} is 0.3 eV larger than that quoted
by Van Deurzen et al.\cite{vd}  This difference is due to different definitions
of the odd L interactions which in turn alters the values of the even L
Slater integrals.\cite{gold1}  Inclusion of these odd L terms improves the
RMS error of the current work by over a factor of 3, not surprising since 3
more fit parameters have been added.\cite{foot}  First principles calculations
of these odd L terms have had mixed success,\cite{mr} so their use is somewhat
dangerous.

Also reported in Fig. 3 and Table 1 are results for $\lambda$=2.6.
This value was chosen to fit the $^3H_4$ - $^3F_2$ splitting in $UPt_3$
observed with high energy neutron scattering.\cite{osb}  This can then be
taken as a prediction for the entire multiplet spectrum in this case.  It
should be remarked, though, that given the itinerancy of the f electrons in
$UPt_3$, such multiplet effects will be very weak.  They will probably be
better observable in $UPd_3$, where the f electrons are known to be localized
and which has an almost identical observed splitting.
Of more interest is the U value, which is necessary in many-body calculations.
The predicted free ion value of 3.3 eV is larger than the experimental
values quoted by Brewer\cite{brewer} of 2.3 and 2.6 eV, although it should be
remarked that these values were extracted from neutral ions with an $f^3$
ground state, and thus this U is expected to be smaller than the one
appropriate
for the more localized $U^{4+}$ ion.  Also, U is predicted here
to be reduced by a factor of two in the metallic environment appropriate for
$UPt_3$ and $UPd_3$.  This U value of 1.6 eV should be a useful input for
many-body calculations in uranium heavy fermion metals.\cite{stein}

In conclusion, a simple method has been proposed for determining the effective
Coulomb integrals needed for many-body calculatons, including the Coulomb U.
This method has been applied to $Pr^{3+}$ and $U^{4+}$ ions and gives
reasonable values for these parameters.  In turn, this method was used to
give a prediction of such values for the case of uranium heavy fermion metals.

\acknowledgments

This work was supported by the U.S. Dept. of Energy,
Basic Energy Sciences, under Contract No. W-31-109-ENG-38.  The author
acknowledges Ray Osborn and Charlotte Froese Fischer for helpful discussions.

\vfill\eject

\begin{table}\caption{Slater integrals (eV) for $Pr^{3+}$, $U^{4+}$, and
$UPt_3$.  In column 2, HF is Hartree-Fock, exp is experiment, and the numbers
refer to values of $\lambda$.  The experimental estimates were taken from
Ref. \protect\cite{brewer,lang} ($Pr$, L=0), Ref. \protect\cite{gold1}
($Pr^{3+}$, L$>$0), Ref. \protect\cite{brewer} ($U$, L=0), and
Ref. \protect\cite{vd} ($U^{4+}$, L$>$0).}
\begin{tabular}{llrrrr}
          &     & $F^0$  & $F^2$  & $F^4$ & $F^6$ \\
\tableline
$Pr^{3+}$ & HF  & 25.722 & 12.227 & 7.670 & 5.517 \\
          & 1.9 &  5.560 &  9.000 & 6.811 & 5.190 \\
          & 2.0 &  5.230 &  8.784 & 6.735 & 5.158 \\
          & exp &5.3,5.5 &  9.090 & 6.927 & 4.756 \\
\tableline
$U^{4+}$  & HF  & 18.796 & 9.482 & 6.202 & 4.552 \\
          & 1.6 &  3.309 & 6.377 & 5.281 & 4.185 \\
          & exp &2.3,2.6 & 6.440 & 5.296 & 3.441 \\
\tableline
$UPt_3$   & 2.6 & 1.645 & 4.421 & 4.336 & 3.724 \\
\end{tabular}
\end{table}

\begin{figure}
\caption{Slater integrals (eV) versus $\lambda$ (Thomas-Fermi wavevector in
a.u.) for the $U^{4+}$ ion.}
\label{fig1}
\end{figure}

\begin{figure}
\caption{Multiplet spectra (eV) relative to the $^3H_4$ ground state of
$Pr^{3+}$ for Hartree-Fock (HF), $\lambda$=2.0 (2.0), experiment
(exp)\protect\cite{gold1}, and configuration-interaction
(CI)\protect\cite{cai}.  The first excited state is $^3H_5$.}
\label{fig2}
\end{figure}

\begin{figure}
\caption{Multiplet spectra (eV) relative to the $^3H_4$ ground state of
$U^{4+}$ for Hartree-Fock (HF), $\lambda$=1.6 (1.6), experiment
(exp)\protect\cite{vd}, and $\lambda$=2.6 (2.6).  The latter $\lambda$
was fit to the observed multiplet splitting in $UPt_3$.\protect\cite{osb}
The first excited state is $^3F_2$ except for Hartee-Fock where it is $^3H_5$.}
\label{fig3}
\end{figure}


\begin{references}
\bibitem{foot} This requires that $\lambda$ be reduced to 1.4 since the odd L
terms are now accounting for some of the correlation.  The fitted odd L terms
are close to the reported values except for the $\beta$ term which is
significantly differently (the $\beta$ term only affects $^1S_0$).
\end{references}
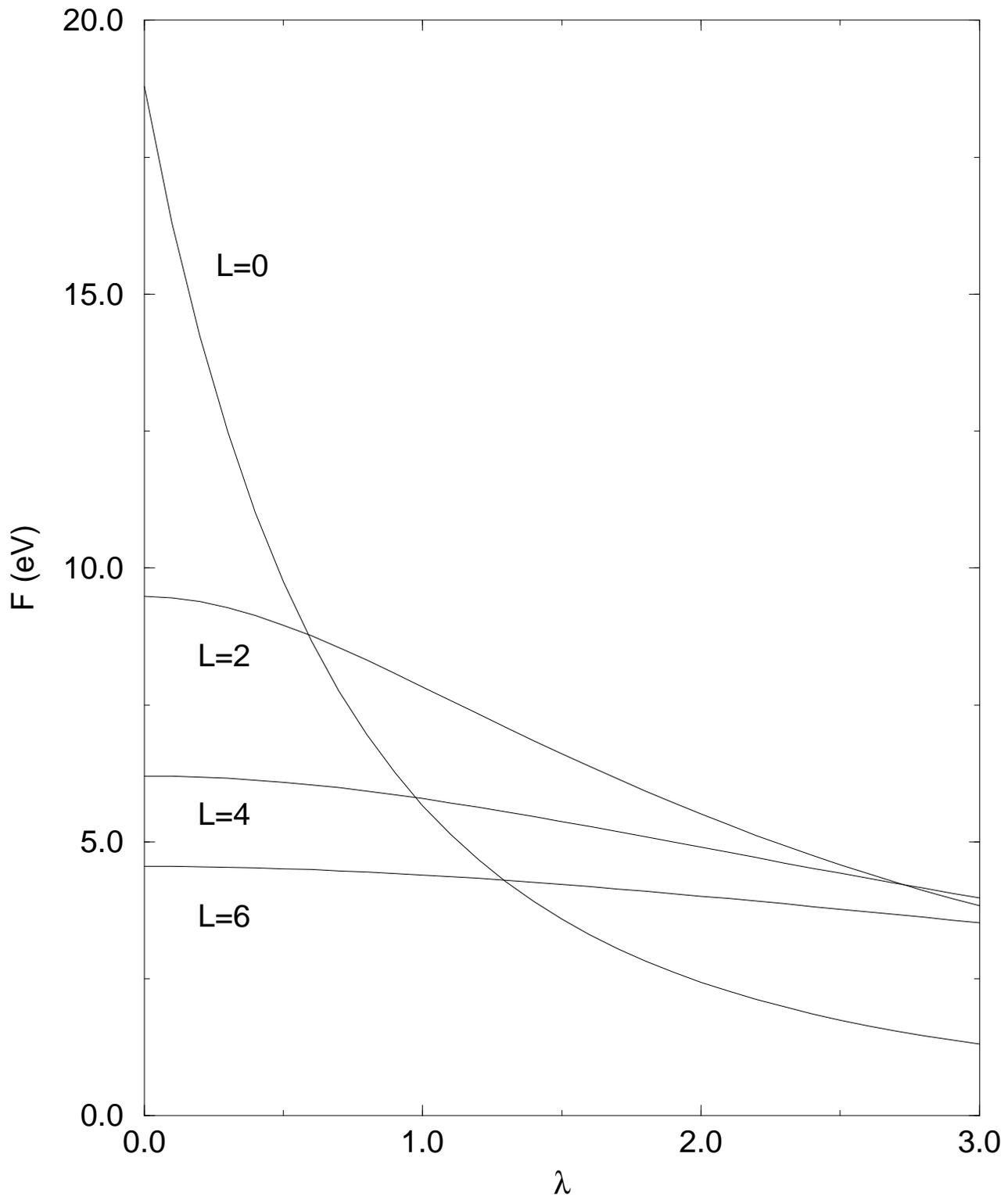
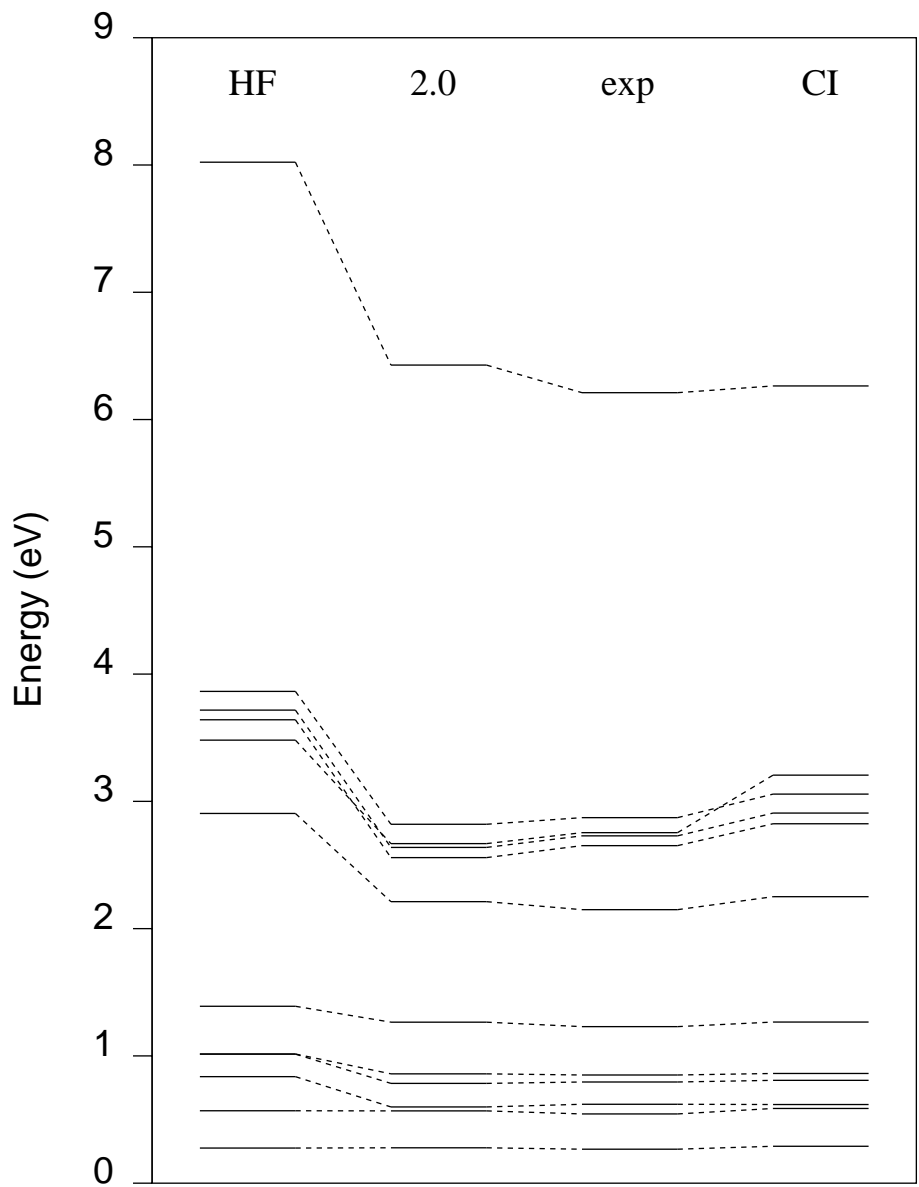
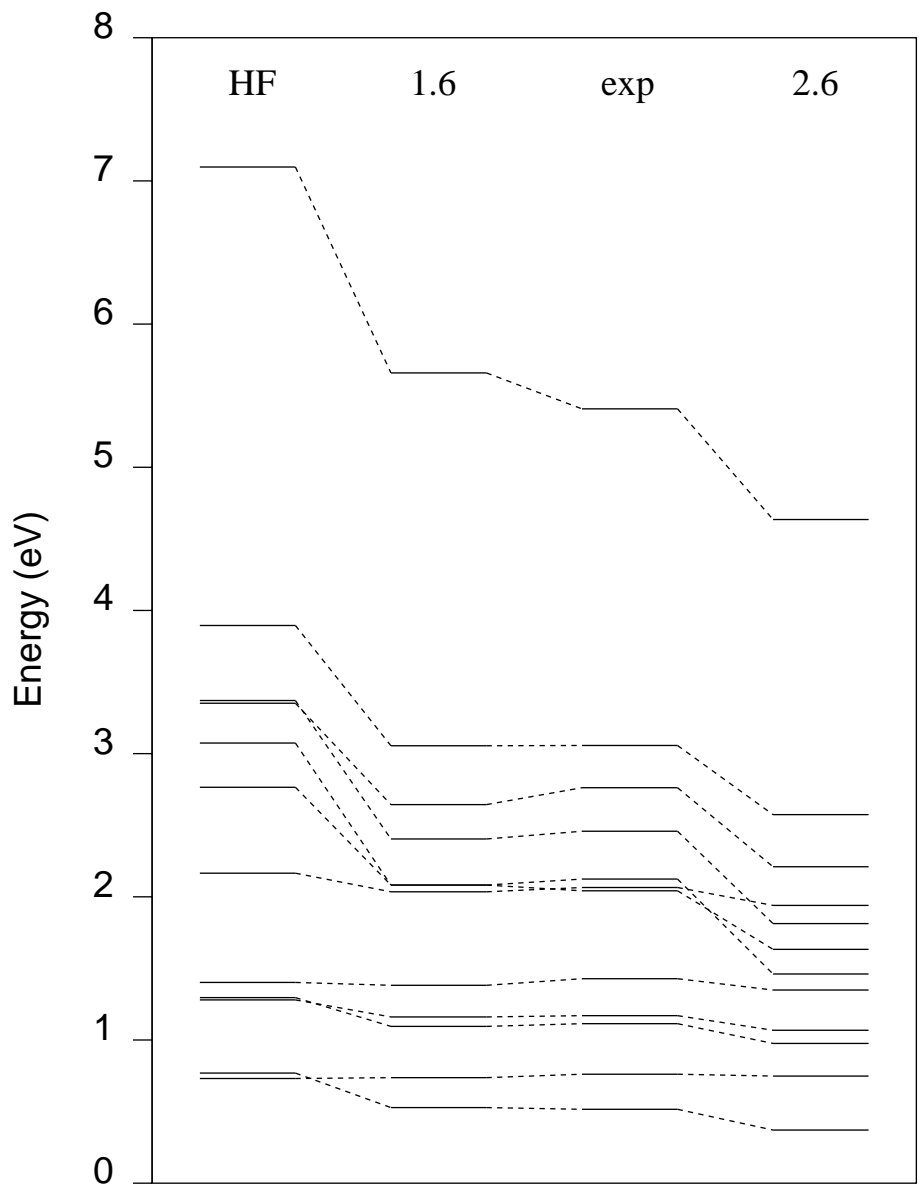
\end{document}